\documentclass[10pt,doublesided, doublecolumn,final]{IEEEtran}
\usepackage{float}
\usepackage{algorithm}
\usepackage[noend]{algorithmic}
\usepackage[geometry]{ifsym}
\usepackage {ifsym}
\usepackage{bbm}
\usepackage{amssymb}
\usepackage{yhmath}
\usepackage{amscd}
\usepackage{dsfont}
\usepackage{amsmath}
\usepackage{epsfig}
\usepackage{graphics}
\usepackage{psfrag}
\usepackage{rotating}
\usepackage{amsmath}
\usepackage{amsfonts}
\usepackage{url}
\usepackage{color}
\usepackage{float}
\usepackage{balance} 

\newtheorem{theorem}{Theorem}
\newtheorem{definition}{Definition}

\newcommand{\bs}{\boldsymbol}

\newcommand{\SNR}{\mathrm{SNR}}

\newcommand{\INR}{\mathrm{INR}}

\newcommand{\sfT}{\textsf{T}}

\newcommand{\Cgicnof}{\mathcal{C}_{\mathrm{GIC-NOF}}}
\newcommand{\agicnof}{\underline{\mathcal{C}}_{\mathrm{G-IC-NOF}}}
\newcommand{\cgicnof}{\overline{\mathcal{C}}_{\mathrm{G-IC-NOF}}}

\begin{document}
\title{Approximate Capacity of the Gaussian Interference Channel with Noisy Channel-Output Feedback}

\author{Victor Quintero, Samir M. Perlaza, I\~naki Esnaola and Jean-Marie Gorce
\thanks{Victor Quintero, Samir M. Perlaza and Jean-Marie Gorce are with the CITI Laboratory, a joint lab between the Institut National de Recherche en Informatique et en Automatique (INRIA), Universit\'{e} de Lyon and Institut National de Sciences Apliqu\'ees (INSA) de Lyon. 6 Av. des Arts 69621 Villeurbanne, France. (victor.quintero-florez@inria.fr).}
\thanks{I\~naki Esnaola is with the University of Sheffield, Dep. of Automatic Control and Systems Engineering, Mappin Street, Sheffield, S1 3JD, United Kingdom.} 
\thanks{Victor Quintero is also with Universidad del Cauca, Popay\'{a}n, Colombia.} 
\thanks{This research was supported in part by the European Commission under Marie Sklodowska-Curie Individual Fellowship No. 659316 (CYBERNETS) and the Administrative Department of Science, Technology and Innovation of Colombia (Colciencias), fellowship No. 617-2013.}
 }

\maketitle

\begin{abstract}  
In this paper, an achievability region and a converse region for the two-user Gaussian interference channel with noisy channel-output feedback (G-IC-NOF)  are presented. The achievability region is obtained using a random coding argument and three well-known techniques: rate splitting, superposition coding and backward decoding. The converse region is obtained using some of the existing perfect-output feedback outer-bounds as well as a set of new outer-bounds that are obtained by using  genie-aided models of the original G-IC-NOF. Finally, it is shown that the achievability region and the converse region approximate the capacity region of the G-IC-NOF to within a constant gap in bits per channel use. 
\end{abstract}

\vspace{-3mm}

\section{Notation}

Throughout this paper, $(\cdot)^+$ denotes the positive part operator, i.e., $(\cdot)^+ = \max(\cdot, 0)$ and $\mathbb{E}_{X}[ \cdot ]$ denotes the expectation with respect to the distribution of the random variable $X$. The logarithm function $\log$ is assumed to be base $2$. 

\vspace{-3mm}

\section{System Model} \label{SectSystMod}

\begin{figure}[t!]
 \centerline{\epsfig{figure=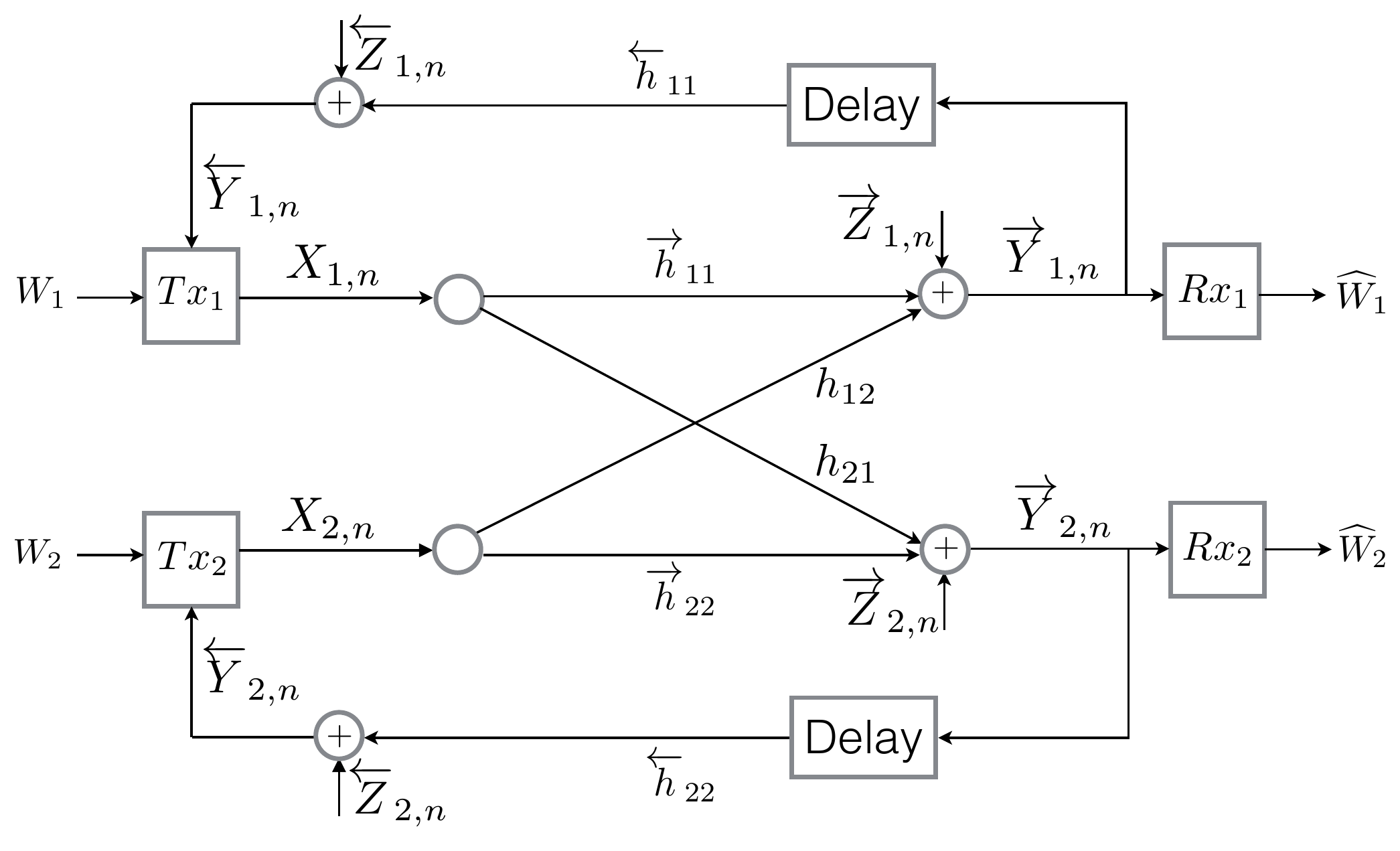,width=0.4\textwidth}}
  \caption{Gaussian interference channel with noisy channel-output feedback at channel use~$n$.}
  \label{Fig:G-IC-NOF}
\end{figure}

Consider the two-user G-IC-NOF in Figure~\ref{Fig:G-IC-NOF}. Transmitter $i$, with $i \in \{1,2\}$, communicates with receiver $i$ subject to the interference produced by transmitter $j$, with $j \in \{1,2\} \backslash \{i\}$. There are two independent and uniformly distributed messages, $W_i \in \mathcal{W}_i$, with $\mathcal{W}_i=\{1, 2,  \ldots, 2^{NR_i}\}$, where $N$ denotes the block-length in channel uses and $R_i$ is the transmission rate in bits per channel use. At each block, transmitter $i$  sends the codeword ${\bs{X}_{i}=\left(X_{i,1}, X_{i,2}, \ldots, X_{i,N}\right)^\sfT \in \mathcal{X}_i^N}$, where $\mathcal{X}_i$ and $\mathcal{X}_i^N$ are respectively the channel-input alphabet and the codebook of transmitter $i$. 

\noindent
The channel coefficient from transmitter $j$ to receiver $i$ is denoted by $h_{ij}$; the channel coefficient from transmitter $i$ to receiver $i$ is denoted by $\overrightarrow{h}_{ii}$; and the channel coefficient from channel-output $i$ to transmitter $i$ is denoted by $\overleftarrow{h}_{ii}$. All channel coefficients are assumed to be non-negative real numbers.
At a given channel use $n \in \{1, 2, \ldots, N\}$, the channel output at receiver $i$ is denoted by $\overrightarrow{Y}_{i,n}$.  
During channel use $n$, the input-output relation of the channel model is given by
\begin{IEEEeqnarray}{lcl}
\label{Eqsignalyif}
\overrightarrow{Y}_{i,n}&=& \overrightarrow{h}_{ii}X_{i,n} + h_{ij}X_{j,n}+\overrightarrow{Z}_{i,n},
\end{IEEEeqnarray}
where $\overrightarrow{Z}_{i,n}$ is a real Gaussian random variable with zero mean and unit variance that represents the noise at the input of receiver $i$.
%
%
Let $d>0$ be the finite feedback delay measured in channel uses. At the end of channel use $n$, transmitter $i$ observes $\overleftarrow{Y}_{i,n}$, which consists of a scaled and noisy version of $\overrightarrow{Y}_{i,n-d}$. More specifically,
\begin{IEEEeqnarray}{rcl}
\label{Eqsignalyib}
\overleftarrow{Y}_{i,n}  &=& 
\begin{cases}
 \overleftarrow{Z}_{i,n} &  \textrm{for } n \! \in \lbrace \! 1, \! 2,  \ldots, d \rbrace  \\ 
\overleftarrow{h}_{ii}\overrightarrow{Y}_{i,n-d} \! + \! \overleftarrow{Z}_{i,n}, \!  &  \textrm{for } n \! \in \lbrace  d \! + \! 1, \! d \! + \! 2, \ldots, \! N \rbrace,
 \end{cases} \quad
\end{IEEEeqnarray}

\noindent
where $\overleftarrow{Z}_{i,n}$ is a real Gaussian random variable with zero mean and unit variance that represents the noise in the feedback link of transmitter-receiver pair  $i$. The random variables $\overrightarrow{Z}_{i,n}$ and $\overleftarrow{Z}_{i,n}$ are independent and identically distributed.
In the following, without loss of generality, the feedback delay is assumed to be one channel use, i.e., $d=1$. 
The encoder of transmitter $i$ is defined by a set of deterministic functions $f_i^{(1)}, \ldots, f_i^{(N)}$, with $f_i^{(1)}:\mathcal{W}_i \rightarrow \mathcal{X}_i$ and for all $n \in \{2, \ldots, N\}$, $f_i^{(n)}:\mathcal{W}_i\times\mathds{R}^{n-1} \rightarrow \mathcal{X}_i$, such that
\begin{subequations}
\label{Eqencod}
\begin{IEEEeqnarray}{lcl}
\label{Eqencodi1}
X_{i,1}&=&f_i^{(1)}\left(W_i\right),  \mbox{ and }  \\
\label{Eqencodit}
X_{i,n}&=&f_i^{(n)}\left(W_i,\overleftarrow{Y}_{i,1},\ldots,\overleftarrow{Y}_{i,n-1}\right).
\end{IEEEeqnarray}
\end{subequations}
The components of the input vector $\bs{X}_{i}$ are real numbers subject to an average power constraint:
\begin{equation}
\label{Eqconstpow}
\frac{1}{N}\sum_{n=1}^{N}\mathbb{E}\left({X_{i,n}}^2\right) \leq 1,
\end{equation}
where the expectation is taken over the joint distribution of the message indexes $W_1$, $W_2$, and the noise terms, i.e., $\overrightarrow{Z}_{1}$, $\overrightarrow{Z}_{2}$, $\overleftarrow{Z}_{1}$, and $\overleftarrow{Z}_{2}$. The dependence of $X_{i,n}$ on $W_1$, $W_2$, and the previously observed noise realizations is due to the effect of feedback as shown in \eqref{Eqsignalyib} and \eqref{Eqencod}. 
%
%

\noindent
Assume that during a given communication, $T$ blocks are transmitted. Hence, the decoder of receiver $i$ is defined by a deterministic function $\psi_i: \mathds{R}_i^{N T} \rightarrow \mathcal{W}_i^{T}$.
At the end of the communication, receiver $i$ uses the vector $\left(\overrightarrow{Y}_{i,1}, \overrightarrow{Y}_{i,2}, \ldots, \overrightarrow{Y}_{i,NT}\right)^\sfT$ to obtain an estimate of the message indices
\begin{IEEEeqnarray}{rcl}
\label{Eqdecoder}
\left(\widehat{W}_i^{(1)}, \widehat{W}_i^{(2)}, \ldots, \widehat{W}_i^{(T)}\right) &=& \psi_i \left(\overrightarrow{Y}_{i,1}, \overrightarrow{Y}_{i,2}, \ldots, \overrightarrow{Y}_{i,NT}\right), \quad
\end{IEEEeqnarray} 
where $\widehat{W}_i^{(t)}$ is an estimate of the message index sent during block $t \in \lbrace 1, 2, \ldots, T \rbrace$.
The decoding error probability in the two-user G-IC-NOF during block $t$ of a codebook of block-length $N$, denoted by $P_{e}^{(t)}(N)$, is given by    
\begin{IEEEeqnarray}{rcl}
\label{EqDecErrorProb}
\! P_{e}^{(t)} \! ( N ) &=&   \max \!  \left(  \textrm{Pr} \! \left[ \! \widehat{W_1}^{(t)} \!  \neq \!  W_1^{(t)} \! \right]  ,  \textrm{Pr} \! \left[ \! \widehat{W_2}^{(t)} \!  \neq \!  W_2^{(t)}  \right]   \right) \! .
\end{IEEEeqnarray}

\noindent
The definition of an achievable rate pair $(R_1,R_2) \in \mathds{R}_+^{2}$ is given below. 
\begin{definition}[Achievable Rate Pairs]\label{DefAchievableRatePairs}\emph{
A rate pair $(R_1,R_2) \in \mathds{R}_+^{2}$ is achievable if there exists at least one pair of codebooks $\mathcal{X}_1^{N}$ and $\mathcal{X}_2^{N}$ with codewords of length $N$, and the corresponding encoding functions $f_1^{(1)},\ldots,f_1^{(N)}$ and $f_2^{(1)},\ldots,f_2^{(N)}$ such that the decoding error probability $P_{e}^{(t)}(N)$ can be made arbitrarily small by letting the block-length $N$ grow to infinity, for all blocks $t \in \lbrace 1, \ldots, T \rbrace$.
 }
\end{definition}

The two-user G-IC-NOF in Figure~\ref{Fig:G-IC-NOF} can be fully described by six parameters: $\overrightarrow{\SNR}_i$, $\overleftarrow{\SNR}_i$, and $\INR_{ij}$, with $i \in \{1,2\}$ and $j \in \{1,2\} \backslash \{i\}$, which are defined as follows:
\begin{IEEEeqnarray}{rcl}
\label{EqSNRifwd}
\overrightarrow{\SNR}_i &=& \overrightarrow{h}_{ii}^2, \\
%
\label{EqINRij}
\INR_{ij}&=& h_{ij}^2 \mbox{ and } \\
\label{EqSNRibwd}
\overleftarrow{\SNR}_i&=&\overleftarrow{h}_{ii}^2\left(\overrightarrow{h}_{ii}^2 + 2\overrightarrow{h}_{ii}h_{ij}+h_{ij}^2+1\right). \quad
\end{IEEEeqnarray}
%
%
%
%

\section{Main Results} \label{SectMainRes}

This section introduces an achievable region (Theorem~\ref{TheoremA-G-IC-NOF})  and a converse region (Theorem~\ref{TheoremC-G-IC-NOF}), denoted by $\agicnof$ and $\cgicnof$ respectively, for the two-user G-IC-NOF with fixed parameters $\overrightarrow{\SNR}_{1}$, $\overrightarrow{\SNR}_{2}$, $\INR_{12}$, $\INR_{21}$, $\overleftarrow{\SNR}_{1}$, and $\overleftarrow{\SNR}_{2}$.
In general, the capacity region of a given multi-user channel is said to be approximated to within a constant gap according to the following definition.

\begin{definition}[Approximation to within $\xi$ units]\label{DefGap}\emph{
A closed and convex set $\mathcal{T}\subset\mathbb{R}_{+}^{m}$ is approximated to within $\xi$ units by the sets $\underline{\mathcal{T}}$ and $\overline{\mathcal{T}}$ if  $\underline{\mathcal{T}} \subseteq \mathcal{T} \subseteq \overline{\mathcal{T}}$ and for all $\bs{t}=(t_1, \ldots, t_m)\in \overline{\mathcal{T}}$ then $((t_1-\xi)^+, \ldots, (t_m-\xi)^+) \in \underline{\mathcal{T}}$.}
\end{definition}

\noindent
Denote by $\Cgicnof$ the capacity region of the 2-user G-IC-NOF.  The achievable region $\agicnof$ and the converse region $\cgicnof$ approximate the capacity region $\Cgicnof$ to within $4.4$ bits per channel use (Theorem~\ref{TheoremGAP-G-IC-NOF}).


\subsection{An Achievable Region for the Two-User G-IC-NOF}
The description of the achievable region $\agicnof$ is presented using the constants $a_{1,i}$; the functions $a_{2,i}:[0,1] \rightarrow \mathds{R}_{+}$,  $a_{l,i}:[0,1]^2\rightarrow \mathds{R}_{+}$, with $l \in \lbrace 3, \ldots, 6 \rbrace$; and $a_{7,i}:[0,1]^3\rightarrow \mathds{R}_{+}$, which are defined as follows, for all $i \in \lbrace 1, 2 \rbrace$, with $j \in \lbrace 1, 2 \rbrace \setminus \lbrace i \rbrace$:

\begin{subequations}
\label{Eq-a}
\begin{IEEEeqnarray}{rcl}
\label{Eq-a1}
a_{1,i}  &=&  \frac{1}{2}\log \left(2+\frac{\overrightarrow{\SNR_{i}}}{\INR_{ji}}\right)-\frac{1}{2}, \\
\label{Eq-a2}
a_{2,i}(\rho) &=& \frac{1}{2}\log \Big(b_{1,i}(\rho)+1\Big)-\frac{1}{2}, \\
\nonumber
a_{3,i}(\rho,\mu) &=& \!  \frac{1}{2}\!\log\! \left(\! \frac{\! \overleftarrow{\SNR}_i \! \Big(b_{2,i}(\rho)+2\Big)+b_{1,i}(1)+1}{\! \overleftarrow{\SNR}_i \! \Big( \! \left(1\!-\!\mu\right) \! b_{2,i}(  \rho  )\!+\! 2 \!\Big)\!+\! b_{1,i}( 1  ) \! + \!1 \!} \! \right) \!, \\ 
\label{Eq-a3}\\
\label{Eq-a4}
a_{4,i}(\rho,\mu) &=& \frac{1}{2}\log \bigg(\Big(1-\mu\Big)b_{2,i}(\rho)+2 \bigg)-\frac{1}{2},\\
\nonumber
a_{5,i}(\rho,\mu) &=& \frac{1}{2}\log \left(2+\frac{\overrightarrow{\SNR}_{i}}{\INR_{ji}}+\Big(1-\mu\Big)b_{2,i}(\rho)\right)-\frac{1}{2},\\
\label{Eq-a5}\\
\nonumber
a_{6,i}(\rho,\mu) &=& \frac{1}{2}\!\log\! \left(\!\frac{\overrightarrow{\SNR}_{i}}{\INR_{ji}}\bigg(\Big(1\!-\!\mu\Big)b_{2,j}(\rho)\!+\!1\bigg)\!+\!2\right)\!-\!\frac{1}{2},\\
\textrm {and } \qquad \quad  & &  \label{Eq-a6}\\
\nonumber
a_{7,i}(\rho,\!\mu_1\!,\!\mu_2\!) &=& \frac{1}{2}\!\log \Bigg(\!\frac{\overrightarrow{\SNR}_{i}}{\INR_{ji}}\bigg(\Big(1\!-\!\mu_i\Big)b_{2,j}(\rho)\!+\!1\bigg) \\
\label{Eq-a7}
& & +\Big(1\!-\!\mu_j\Big)b_{2,i}(\rho)+2\Bigg)\!-\!\frac{1}{2},
\end{IEEEeqnarray}
\end{subequations}
where the functions $b_{l,i}:[0,1]\rightarrow \mathds{R}_{+}$, with $(l,i) \in \lbrace1, 2 \rbrace^2$ are defined as follows: 
\begin{subequations}
\label{Eqfnts}
\begin{IEEEeqnarray}{rcl}
\label{Eqb1i}
b_{1,i}(\rho)&=&\overrightarrow{\SNR}_{i}+2\rho\sqrt{\overrightarrow{\SNR}_{i}\INR_{ij}}+\INR_{ij} \mbox{ and } \\
\label{Eqb5i}
b_{2,i}(\rho)&=&\Big(1-\rho\Big)\INR_{ij}-1,
\end{IEEEeqnarray}
\end{subequations}
with $j \in \lbrace 1, 2 \rbrace \setminus \lbrace i \rbrace$.

\noindent
Note that the functions in \eqref{Eq-a} and \eqref{Eqfnts} depend on $\overrightarrow{\SNR}_{1}$, $\overrightarrow{\SNR}_{2}$, $\INR_{12}$, $\INR_{21}$, $\overleftarrow{\SNR}_{1}$, and $\overleftarrow{\SNR}_{2}$, however as these parameters are fixed in this analysis, this dependence is not emphasized in the definition of these functions. Finally, using this notation, Theorem~\ref{TheoremA-G-IC-NOF} is presented on the next page.
\begin{figure*}[t]
\begin{theorem} \label{TheoremA-G-IC-NOF} \emph{
The capacity region $\Cgicnof$ contains the region $\agicnof$ given by the closure of the set of all possible non-negative achievable rate pairs $(R_1,R_2)$ that satisfy
\begin{subequations}
\label{EqRa-G-IC-NOF}
\begin{IEEEeqnarray}{rcl}
\label{EqR1a-G-IC-NOF}
R_{1}  & \leqslant & \min\Big(a_{2,1}(\rho),a_{6,1}(\rho,\mu_1)+a_{3,2}(\rho,\mu_1), a_{1,1}+a_{3,2}(\rho,\mu_1)+a_{4,2}(\rho,\mu_1)\Big),  \\ 
\label{EqR2a-G-IC-NOF}
R_{2}   & \leqslant & \min\Big(a_{2,2}(\rho),a_{3,1}(\rho,\mu_2)+a_{6,2}(\rho,\mu_2), a_{3,1}(\rho,\mu_2)+a_{4,1}(\rho,\mu_2)+a_{1,2}\Big),   \\
\nonumber
R_{1}+R_{2}  & \leqslant & \min\Big(a_{2,1}(\rho)+a_{1,2}, a_{1,1}+a_{2,2}(\rho), a_{3,1}(\rho,\mu_2)+a_{1,1}+a_{3,2}(\rho,\mu_1)+a_{7,2}(\rho,\mu_1,\mu_2), \\
\label{EqR1+R2a-G-IC-NOF}
& & a_{3,1}(\rho,\mu_2)+a_{5,1}(\rho,\mu_2)+a_{3,2}(\rho,\mu_1)+a_{5,2}(\rho,\mu_1), a_{3,1}(\rho,\mu_2)+a_{7,1}(\rho,\mu_1,\mu_2)+a_{3,2}(\rho,\mu_1)+a_{1,2}\Big),  \\
\label{Eq2R1+R2a-G-IC-NOF}
2R_{1}+R_{2}  & \leqslant & \min\Big(a_{2,1}(\rho)+a_{1,1}+a_{3,2}(\rho,\mu_1)+a_{7,2}(\rho,\mu_1,\mu_2),  \\
\nonumber
& &  a_{3,1}(\rho,\mu_2)+a_{1,1}+a_{7,1}(\rho,\mu_1,\mu_2)+2a_{3,2}(\rho,\mu_1)+a_{5,2}(\rho,\mu_1), a_{2,1}(\rho)+a_{1,1}+a_{3,2}(\rho,\mu_1)+a_{5,2}(\rho,\mu_1)\Big), \\
\nonumber
R_{1}+2R_{2}  & \leqslant & \min\Big(a_{3,1}(\rho,\mu_2)+a_{5,1}(\rho,\mu_2)+a_{2,2}(\rho)+a_{1,2}, a_{3,1}(\rho,\mu_2)+a_{7,1}(\rho,\mu_1,\mu_2)+a_{2,2}(\rho)+a_{1,2}, \\
\label{EqR1+2R2a-G-IC-NOF}
& &  2a_{3,1}(\rho,\mu_2)+a_{5,1}(\rho,\mu_2)+a_{3,2}(\rho,\mu_1)+a_{1,2}+a_{7,2}(\rho,\mu_1,\mu_2)\Big),
\end{IEEEeqnarray}
\end{subequations}
with $\left(\rho, \mu_1, \mu_2\right) \in \left[0,\left(1-\max\left(\frac{1}{\INR_{12}},\frac{1}{\INR_{21}}\right) \right)^+\right]\times[0,1]\times[0,1]$.
}
\end{theorem}
\vspace{-5mm}
\end{figure*}

\begin{IEEEproof}
The proof of Theorem~\ref{TheoremA-G-IC-NOF} is presented in \cite{QPEG-TR-2016-2}.
\end{IEEEproof}
\subsection{Comments on the Achievability}
The achievable region is obtained using a random coding argument and combining three classical tools: rate splitting, superposition coding, and backward decoding. This coding scheme is described in \cite{QPEG-TR-2016-2} and it is specially designed for the two-user IC-NOF. Consequently, only the strictly needed number of superposition code-layers is used.  Other achievable schemes, as reported in \cite{SyQuoc-TIT-2015}, can also be obtained as special cases of the more general scheme presented in \cite{Tuninetti-ISIT-2007}. However, in this more general case, the resulting code for the IC-NOF contains a handful of unnecessary superposing code-layers, which complicates the error probability analysis.     
\subsection{A Converse Region for the Two-User G-IC-NOF}

The description of the converse region $\cgicnof$ is determined by the ratios  $\frac{\INR_{ij}}{\overrightarrow{\SNR}_{j}}$, and $\frac{\INR_{ji}}{\overrightarrow{\SNR}_{j}}$, for all $i \in \lbrace 1, 2 \rbrace$, with $j \in \lbrace 1, 2 \rbrace\setminus\lbrace i \rbrace$. All relevant scenarios regarding these ratios are described by two events denoted by $S_{l_{1},1}$ and $S_{l_{2},2}$, where $(l_{1},l_{2}) \in \lbrace 1, \ldots, 5 \rbrace^2$. The events are defined as follows:
 \begin{subequations}
\label{EqSi}
\begin{IEEEeqnarray}{rcl}
\label{EqS1i}  
 S_{1,i}&: \quad & \overrightarrow{\SNR}_{j} < \min\left(\INR_{ij},\INR_{ji}\right), \\ 
\label{EqS2i}
 S_{2,i}&: \quad & \INR_{ji} \leqslant  \overrightarrow{\SNR}_{j} < \INR_{ij},\\
 \label{EqS3i}
 S_{3,i}&: \quad & \INR_{ij} \leqslant \overrightarrow{\SNR}_{j} < \INR_{ji}, \\
 \label{EqS4i}
 S_{4,i}&: \quad & \max\left(\INR_{ij}, \INR_{ji}\right) \leqslant \overrightarrow{\SNR}_{j} < \INR_{ij}\INR_{ji}, \qquad \\
 \label{EqS5i}
 S_{5,i}&: \quad & \overrightarrow{\SNR}_{j} \geqslant \max\left(\INR_{ij}, \INR_{ji}, \INR_{ij}\INR_{ji}\right).
\end{IEEEeqnarray}
\end{subequations}
Note that for all $i \in \lbrace 1, 2 \rbrace$, the events $S_{1,i}$, $S_{2,i} $, $S_{3,i}$, $S_{4,i}$, and $S_{5,i}$ are mutually exclusive. This observation shows that given any $4$-tuple $(\overrightarrow{\SNR}_{1}, \overrightarrow{\SNR}_{2}, \INR_{12}, \INR_{21})$, there always exists one and only one pair of events  $(S_{l_{1},1}, S_{l_{2},2})$, with $(l_{1},l_{2}) \in \lbrace 1, \ldots, 5 \rbrace^2$, that identifies a unique scenario. Note also that the pairs of events $(S_{2,1}, S_{2,2})$ and $(S_{3,1}, S_{3,2})$ are not feasible. In view of this, twenty-three different scenarios can be identified using the events in \eqref{EqSi}.
Once the exact scenario is identified, the converse region is described using the functions $\kappa_{l,i}: [0,1]\rightarrow \mathds{R}_{+}$, with $(l,i) \in \lbrace 1, \ldots, 3\rbrace\times \lbrace 1, 2\rbrace$; $\kappa_{l}: [0,1]\rightarrow \mathds{R}_{+}$, with $l \in \lbrace 4, 5 \rbrace$; $\kappa_{6,l}: [0,1]\rightarrow \mathds{R}_{+}$, with $l \in \lbrace 1, \ldots, 4\rbrace$; and $\kappa_{7,i,l}:[0,1]\rightarrow \mathds{R}_{+}$, with $(i,l) \in \lbrace 1, 2 \rbrace^2$. These functions are defined as follows for all $i \in \lbrace 1, 2 \rbrace$, with $j \in \lbrace 1, 2 \rbrace\setminus\lbrace i \rbrace$:
\begin{subequations}
\label{Eqconv}
\begin{IEEEeqnarray}{rcl}
\label{Eqconv1}
\kappa_{1,i}(\rho)  & = & \frac{1}{2}\log \Big(b_{1,i}(\rho)+1\Big), \\ 
\label{Eqconv2}
\kappa_{2,i}(\rho)  & = & \frac{1}{2}\log \Big(1+b_{5,j}(\rho)\Big) \!+\! \frac{1}{2}\log \Bigg(1\!+\! \frac{b_{4,i}(\rho)}{1+b_{5,j}(\rho)}\Bigg)\!, \, \qquad \\
\nonumber
\kappa_{3,i}(\rho)  & = & \frac{1}{2} \! \log \! \left(\frac{\overleftarrow{\SNR}_j\bigg(b_{4,i}(\rho)+b_{5,j}(\rho)+1\bigg)}{\! \bigg(b_{1,j}(1) \! + \! 1\bigg)\bigg(b_{4,i}(\rho) \! + 1 \! \bigg)} \! + \! 1 \! \right) \quad\\
\label{Eqconv3}
& &+\frac{1}{2}\log\Big(b_{4,i}(\rho)+1\Big),
\end{IEEEeqnarray}
\vspace{-4mm}
\begin{IEEEeqnarray}{rcl}
\label{Eqconv4}
\kappa_{4}(\rho)  & = & \frac{1}{2}\log \Bigg(1+\frac{b_{4,1}(\rho)}{1+b_{5,2}(\rho)}\Bigg) \! + \! \frac{1}{2}\log \Big(b_{1,2}(\rho)+1\Big)\!, \qquad
\end{IEEEeqnarray}
\begin{IEEEeqnarray}{rcl}
\label{Eqconv5}
\kappa_{5}(\rho)  & = & \frac{1}{2}\log \Bigg(1 \! + \! \frac{b_{4,2}(\rho)}{1 \!+ \! b_{5,1}(\rho)}\Bigg) \! + \! \frac{1}{2}\log \Big(b_{1,1}(\rho) \! + \! 1\Big),\qquad
\end{IEEEeqnarray}
\vspace{-4mm}
\begin{IEEEeqnarray}{rcl}
\label{Eqk6} 
\kappa_{6}(\rho) &=& \begin{cases} \kappa_{6,1}(\rho) & \textrm{ if } (S_{1,2} \lor S_{2,2} \lor S_{5,2} )  \\
                                                                      & \quad \land (S_{1,1} \lor S_{2,1} \lor S_{5,1} )\\
                                                \kappa_{6,2} (\rho) & \textrm{if }(S_{1,2} \lor S_{2,2} \lor S_{5,2} ) \\
                                                                      & \quad \land (S_{3,1} \lor S_{4,1} )\\
                                                \kappa_{6,3}(\rho) & \textrm{if } (S_{3,2} \lor S_{4,2}) \\
                                                                      & \quad \land (S_{1,1} \lor S_{2,1} \lor S_{5,1})\\
                                                \kappa_{6,4}(\rho) & \textrm{if } (S_{3,2} \lor S_{4,2} ) \land ( S_{3,1} \lor S_{4,1})
                                           \end{cases} \qquad \qquad
\end{IEEEeqnarray}
\begin{IEEEeqnarray}{rcl}
\label{Eqk77}
\kappa_{7,i}(\rho) &=& \begin{cases} \kappa_{7,i,1}(\rho) & \textrm{if } (S_{1,i} \lor S_{2,i} \lor S_{5,i})\\
                                                  \kappa_{7,i,2}(\rho) & \textrm{if } (S_{3,i} \lor S_{4,i})
                                                \end{cases}\qquad\qquad\qquad\qquad
\end{IEEEeqnarray}
\end{subequations}
\begin{figure*}[t!]
 \centerline{\epsfig{figure=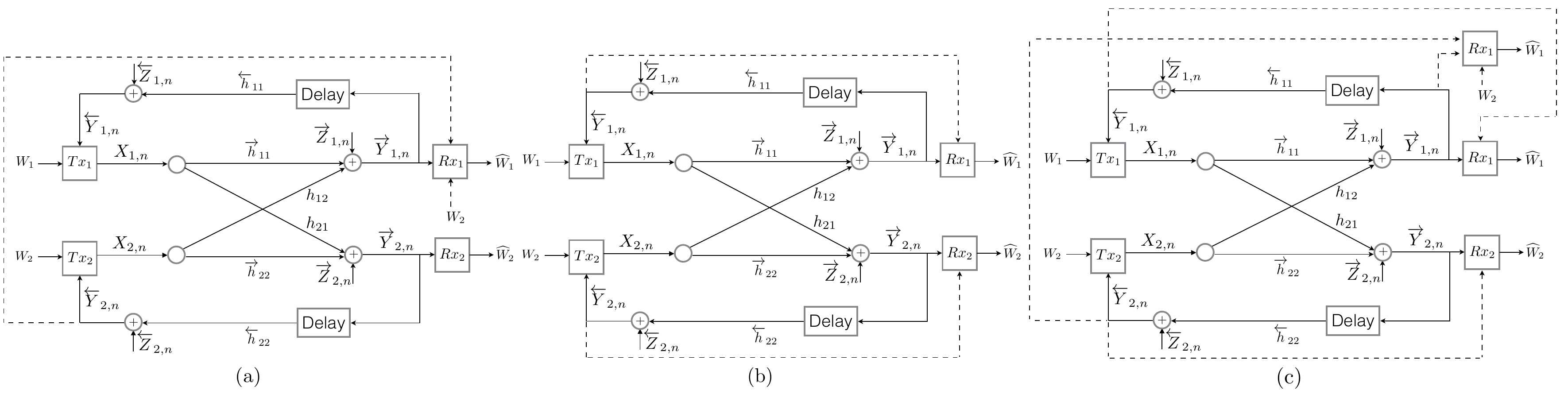,width=1.0\textwidth}}
  \caption{Genie-Aided G-IC-NOF models for channel use~$n$. $(a)$ Model used to calculate the outer-bound on $R_1$; $(b)$ Model used to calculate the outer-bound on $R_1+R_2$; and  $(c)$ Model used to calculate the outer-bound on $2 R_1+R_2$}
  \label{Fig:G-IC-NOF-Conv}
\vspace{-6mm}
\end{figure*}
where
\begin{subequations}
\label{Eqconv6}
\begin{IEEEeqnarray}{rcl}
\nonumber
\kappa_{6,1}&(&\rho) \! = \! \frac{1}{2} \! \log \! \Big(b_{1,1}(\rho) \! + \! b_{5,1}(\rho)\INR_{21}\Big) \! - \! \frac{1}{2}\log\Big(1 \! + \! \INR_{12}\Big) \\
\nonumber
& & +\frac{1}{2}\log\left(1+\frac{b_{5,2}(\rho)\overleftarrow{\SNR}_2}{b_{1,2}(1)+1}\right)\\
\nonumber
& & +\frac{1}{2}\log\Big(b_{1,2}(\rho)+b_{5,1}(\rho)\INR_{21}\Big) -\frac{1}{2}\log\Big(1 \! + \! \INR_{21}\Big) \! \\
\label{Eqconv61}
& &  + \! \frac{1}{2}\log\left(\! 1 \! + \! \frac{b_{5,1}(\rho)\overleftarrow{\SNR}_1}{b_{1,1}(1)+1}\right)+\log(2 \pi e), \\
\nonumber   
\kappa_{6,2}&(&\rho) \! = \!   \frac{1}{2}\log\left(b_{6,2}(\rho)+\frac{b_{5,1}(\rho)\INR_{21}}{\overrightarrow{\SNR}_2}\Big(\overrightarrow{\SNR}_2+b_{3,2}\Big)\right)\\
\nonumber
& & -  \frac{1}{2}\log\Big(1 \!+ \! \INR_{12}\Big) +\frac{1}{2}\log\left(1+\frac{b_{5,1}(\rho)\overleftarrow{\SNR}_1}{b_{1,1}(1)+1}\right) \\
\nonumber
& & + \frac{1}{2} \! \log\Big(b_{1,1}(\rho) \! + \! b_{5,1}(\rho)\INR_{21}\Big) \! -\frac{1}{2}\log\Big(1+\INR_{21}\Big)\\
\nonumber
& & +\frac{1}{2}\log\Bigg(1+\frac{b_{5,2}(\rho)}{\overrightarrow{\SNR}_2}\left(\INR_{12}+\frac{b_{3,2} \overleftarrow{\SNR}_2}{b_{1,2}(1)+1}\right)\Bigg)\\
\label{Eqconv62}
& & -\frac{1}{2}\log\left(1+\frac{b_{5,1}(\rho)\INR_{21}}{\overrightarrow{\SNR}_2}\right)+\log(2 \pi e),
\end{IEEEeqnarray}
\begin{IEEEeqnarray}{rcl}
\nonumber
\kappa_{6,3}&(&\rho) \! = \! \frac{1}{2}\log\Bigg(b_{6,1}(\rho)+\frac{b_{5,1}(\rho)\INR_{21}}{\overrightarrow{\SNR}_1}\Big(\overrightarrow{\SNR}_1+b_{3,1}\Big)\Bigg)\\
\nonumber
& & -\frac{1}{2}\log\Big(1+\INR_{12}\Big)+\frac{1}{2}\log\left(1+\frac{b_{5,2}(\rho)\overleftarrow{\SNR}_2}{b_{1,2}(1)+1}\right)  \\
\nonumber
& & +\frac{1}{2}\log\Big(b_{1,2}(\rho) \! + \! b_{5,1}(\rho)\INR_{21}\Big) \! - \! \frac{1}{2}\log\Big(1 \! + \! \INR_{21}\Big)\\
\nonumber
& & +\frac{1}{2}\log\Bigg(1+\frac{b_{5,1}(\rho)}{\overrightarrow{\SNR}_1}\left(\INR_{21}+\frac{b_{3,1} \overleftarrow{\SNR}_1}{b_{1,1}(1)+1}\right)\Bigg)\\
\label{Eqconv63}
& & -\frac{1}{2}\log\left(1+\frac{b_{5,1}(\rho)\INR_{21}}{\overrightarrow{\SNR}_1}\right) +\log(2 \pi e), \\
\nonumber
\kappa_{6,4}&(&\rho)=\frac{1}{2}\log\left(b_{6,1}(\rho)+\frac{b_{5,1}(\rho)\INR_{21}}{\overrightarrow{\SNR}_1}\Big(\overrightarrow{\SNR}_1+b_{3,1}\Big)\right)\\
\nonumber
& & -\frac{1}{2}\log\Big(1+\INR_{12}\Big)-\frac{1}{2}\log\Big(1+\INR_{21}\Big)\\
\nonumber
& & +\frac{1}{2}\log\left(1+\frac{b_{5,2}(\rho)}{\overrightarrow{\SNR}_2}\left(\INR_{12}+\frac{b_{3,2}\overleftarrow{\SNR}_2}{b_{1,2}(1)+1}\right)\right)\\
\nonumber
& & -\frac{1}{2}\log\left(1+\frac{b_{5,1}(\rho)\INR_{21}}{\overrightarrow{\SNR}_2}\right)\\
\nonumber
& & -\frac{1}{2}\log\left(1+\frac{b_{5,1}(\rho)\INR_{21}}{\overrightarrow{\SNR}_1}\right)\\
\nonumber
& & +\frac{1}{2}\log\left(b_{6,2}(\rho)+\frac{b_{5,1}(\rho)\INR_{21}}{\overrightarrow{\SNR}_2}\Big(\overrightarrow{\SNR}_2+b_{3,2}\Big)\right)\\
\nonumber
& & +\frac{1}{2}\log\Bigg(1+\frac{b_{5,1}(\rho)}{\overrightarrow{\SNR}_1}\left(\INR_{21}+\frac{b_{3,1} \overleftarrow{\SNR}_1}{b_{1,1}(1)+1}\right)\Bigg)\\
\label{Eqconv64}
& & +\log(2 \pi e),
\end{IEEEeqnarray}
\end{subequations}
and 
\vspace{-2mm}
\begin{subequations}
\label{Eqconv7i}
\begin{IEEEeqnarray}{rcl}
\nonumber
\kappa_{7,i,1}&(&\rho) =\frac{1}{2}\log\Big(b_{1,i}(\rho)+1\Big)-\frac{1}{2}\log\Big(1+\INR_{ij}\Big) \\
\nonumber
& & +\frac{1}{2}\log\left(1+\frac{b_{5,j}(\rho)\overleftarrow{\SNR}_j}{b_{1,j}(1)+1}\right)\\
\nonumber
& & +\frac{1}{2}\log\Big(b_{1,j}(\rho)+b_{5,i}(\rho)\INR_{ji}\Big)\\
\nonumber
& & +\frac{1}{2}\log\Big(1 \! + \! b_{4,i}(\rho) \!+ \! b_{5,j}(\rho) \Big) \! - \! \frac{1}{2}\log\Big(1 \! + \! b_{5,j}(\rho)\Big)\\
\label{Eqconv7i1}
& & +2\log(2 \pi e), 
\end{IEEEeqnarray}
\vspace{-4mm}
\begin{IEEEeqnarray}{rcl}
\nonumber
\kappa_{7,i,2}&(&\rho) = \frac{1}{2}\log\Big(b_{1,i}(\rho)+1\Big)-\frac{1}{2}\log\Big(1+\INR_{ij}\Big)\\
\nonumber
& & -\frac{1}{2}\log\Big(1+b_{5,j}(\rho)\Big)+\frac{1}{2}\log\Big(1+b_{4,i}(\rho)+b_{5,j}(\rho) \Big) \quad\\
\nonumber
& & +\frac{1}{2}\log\Bigg(1+\Big(1-\rho^2\Big)\frac{\INR_{ji}}{\overrightarrow{\SNR}_j}\Bigg(\INR_{ij}+\\
\nonumber
& & \frac{b_{3,j}\overleftarrow{\SNR}_j}{b_{1,j}(1)+1}\Bigg)\Bigg)-\frac{1}{2}\log\left(1+\frac{b_{5,i}(\rho)\INR_{ji}}{\overrightarrow{\SNR}_j}\right)\\
\nonumber
& & +\frac{1}{2}\!\log\!\left(\!b_{6,j}(\rho)\!+\!\frac{b_{5,i}(\rho)\INR_{ji}}{\overrightarrow{\SNR}_j}\Big(\overrightarrow{\SNR}_j+b_{3,j}\Big)\right)\\
\label{Eqconv7i2}
& & +2\log(2 \pi e), \quad
\end{IEEEeqnarray}
\end{subequations}
where the functions $b_{l,i}$, with $(l,i) \in \lbrace1, 2 \rbrace^2$ are defined in \eqref{Eqfnts}; $b_{3,i}$ are constants; and the functions $b_{l,i}:[0,1]\rightarrow \mathds{R}_{+}$, with $(l,i) \in \lbrace 4, 5, 6 \rbrace\times\lbrace1, 2 \rbrace$ are defined as follows, with $j \in \lbrace 1, 2 \rbrace \setminus \lbrace i \rbrace$:
\begin{subequations}
\label{Eqfnts2}
\begin{IEEEeqnarray}{rcl}
\label{Eqb2i}
b_{3,i}&=&\overrightarrow{\SNR}_i-2\sqrt{\overrightarrow{\SNR}_i\INR_{ji}}+\INR_{ji}, \\
\label{Eqb3i}
b_{4,i}(\rho)&=&\Big(1-\rho^2\Big)\overrightarrow{\SNR}_{i}, \\
\label{Eqb4i}
b_{5,i}(\rho)&=&\Big(1-\rho^2\Big)\INR_{ij},\\
\nonumber
b_{6,i}(\rho)&=&\overrightarrow{\SNR}_i+\INR_{ij}+2\rho\sqrt{\INR_{ij}}\left(\sqrt{\overrightarrow{\SNR}_i}-\sqrt{\INR_{ji}}\right)\\
\label{Eqb6i}
& & +\frac{\INR_{ij}\sqrt{\INR_{ji}}}{\overrightarrow{\SNR}_i} \left(\sqrt{\INR_{ji}}-2\sqrt{\overrightarrow{\SNR}_i}\right).
\end{IEEEeqnarray}
\end{subequations}

\noindent
Note that the functions in \eqref{Eqconv}, \eqref{Eqconv6}, \eqref{Eqconv7i} and \eqref{Eqfnts2} depend on $\overrightarrow{\SNR}_{1}$, $\overrightarrow{\SNR}_{2}$, $\INR_{12}$, $\INR_{21}$, $\overleftarrow{\SNR}_{1}$, and $\overleftarrow{\SNR}_{2}$. However, these parameters are fixed in this analysis, and therefore, this dependence is not emphasized in the definition of these functions.
Finally, using this notation, Theorem~\ref{TheoremC-G-IC-NOF} is presented below.
\begin{theorem} \label{TheoremC-G-IC-NOF} \emph{
The capacity region $\Cgicnof$ is contained within the region $\cgicnof$ given by the closure of the set of non-negative rate pairs  $(R_1,R_2)$ that for all $i \in \lbrace 1, 2 \rbrace$, with $j\in\lbrace 1, 2 \rbrace\setminus\lbrace i \rbrace$ satisfy:
\begin{subequations}
\label{EqRic-G-IC-NOF}
\begin{IEEEeqnarray}{rcl}
\label{EqRic-12-G-IC-NOF}
R_{i}  & \leqslant & \min\left(\kappa_{1,i}(\rho), \kappa_{2,i}(\rho)\right), \\ 
\label{EqRic-3-G-IC-NOF}
R_{i}  & \leqslant & \kappa_{3,i}(\rho), \\
\label{EqR1+R2c-12-G-IC-NOF}
R_{1}+R_{2}  & \leqslant & \min\left(\kappa_{4}(\rho), \kappa_{5}(\rho)\right),\\
\label{EqR1+R2c-3g-G-IC-NOF}
R_{1}+R_{2}  & \leqslant &\kappa_{6}(\rho),\\
\label{Eq2Ri+Rjc-g-G-IC-NOF}
2R_i+R_j&\leqslant&  \kappa_{7,i}(\rho),
\end{IEEEeqnarray}
\end{subequations}
with $\rho \in [0,1]$.
}
\end{theorem} 
\begin{IEEEproof}
The proof of Theorem~\ref{TheoremC-G-IC-NOF} is presented in \cite{QPEG-TR-2016-2}.
\end{IEEEproof}

\subsection{Comments on the Converse Region}

The outer bounds \eqref{EqRic-12-G-IC-NOF} and \eqref{EqR1+R2c-12-G-IC-NOF} correspond to the outer bounds for the case of perfect channel-output feedback \cite{Suh-TIT-2011}. 
The bounds \eqref{EqRic-3-G-IC-NOF}, \eqref{EqR1+R2c-3g-G-IC-NOF} and \eqref{Eq2Ri+Rjc-g-G-IC-NOF} correspond to new outer bounds that generalize those presented in \cite{SyQuoc-TIT-2015} for the two-user symmetric G-IC-NOF. These new outer-bounds were obtained using the genie-aided models shown in Figure~\ref{Fig:G-IC-NOF-Conv}.

\subsection{A Gap Between the Achievable Region and the Converse Region}
\balance

\begin{figure}[t]
\centerline{\epsfig{figure=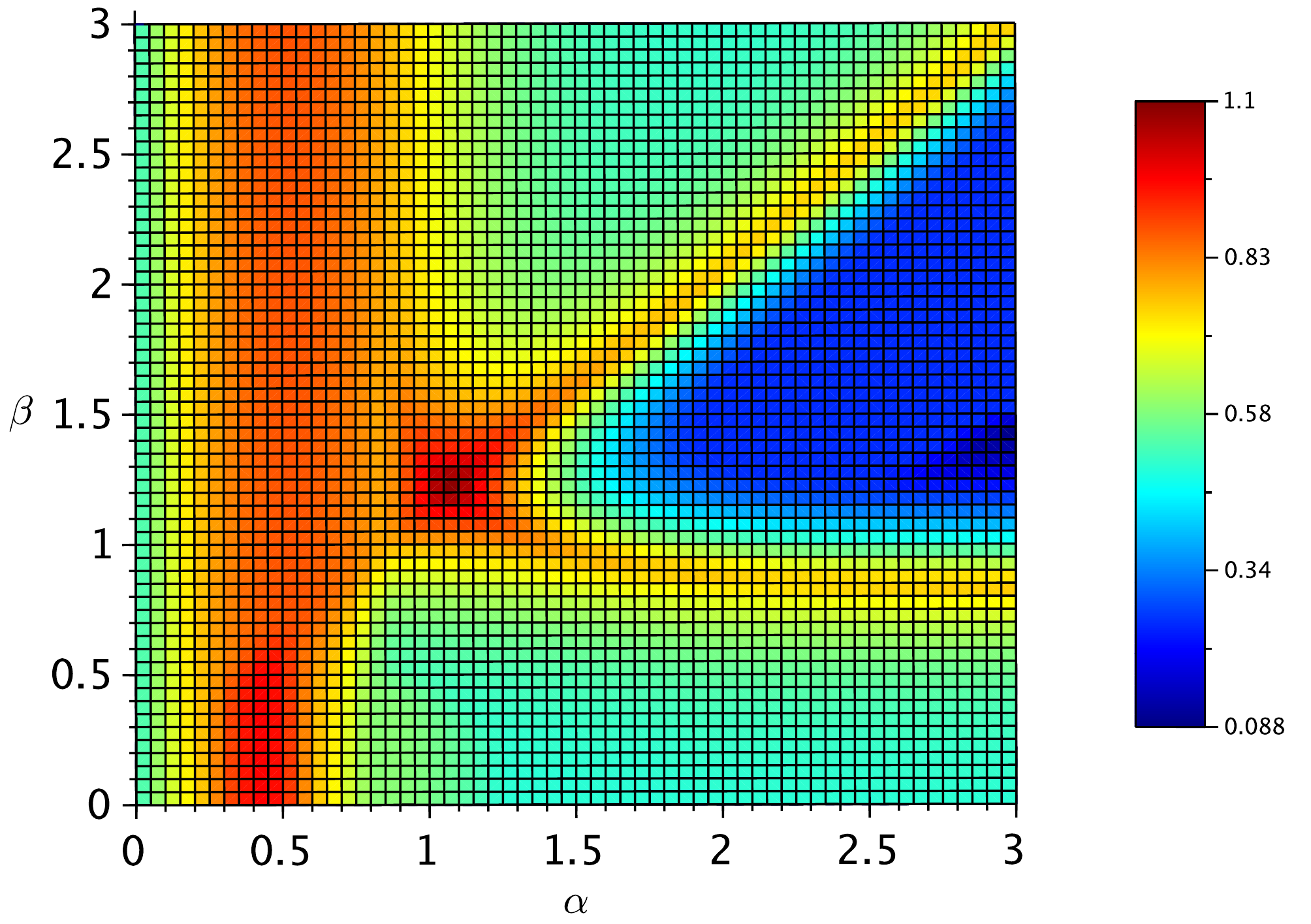,width=0.5\textwidth}}
\caption{Gap between the converse region $\cgicnof$ and the achievable region $\agicnof$ of the two-user G-IC-NOF, under symmetric channel conditions, i.e., $\protect\overrightarrow{\SNR}_1=\protect\overrightarrow{\SNR}_2=\protect\overrightarrow{\SNR}$, $\INR_{12}=\INR_{21}=\INR$, and $\protect\overleftarrow{\SNR}_1=\protect\overleftarrow{\SNR}_2=\protect\overleftarrow{\SNR}$, as a function of $\alpha=\frac{\log \INR}{\log{\protect\overrightarrow{\SNR}}}$ and $\beta=\frac{\log\protect\overleftarrow{\SNR}}{\log\protect\overrightarrow{\SNR}}$.}
\label{FigGapGICNOF}
\end{figure}

Theorem~\ref{TheoremGAP-G-IC-NOF} describes the gap between the achievable region $\agicnof$ and the converse region $\cgicnof$ using the approximation  notion described in Definition~\ref{DefGap}.

\begin{theorem} \label{TheoremGAP-G-IC-NOF} \emph{The capacity region of the two-user G-IC-NOF is approximated to within $4.4$ bits per channel use by the achievable region $\agicnof$ and the converse region $\cgicnof$.
}
\end{theorem} 

\begin{IEEEproof}
The proof of Theorem~\ref{TheoremGAP-G-IC-NOF} is presented in \cite{QPEG-TR-2016-2}.
\end{IEEEproof}

\noindent
The  gap, denoted by $\delta$, between the sets $\cgicnof$ and $\agicnof$ can be approximated (Definition 2) as follows: 
\begin{IEEEeqnarray} {ll}
\label{Eqdelta}
\delta&\leqslant\max\left(\delta_{R_1},\delta_{R_2},\frac{\delta_{2R}}{2},\frac{\delta_{3R_1}}{3}, \frac{\delta_{3R_2}}{3}\right),
\end{IEEEeqnarray}      
where
\begin{subequations}
\label{EqDeltas}
\begin{IEEEeqnarray} {ll}
\nonumber
\delta_{R_1}& \triangleq \min\Big(\kappa_{1,1}(\rho),\kappa_{2,1}(\rho),\kappa_{3,1}(\rho)\Big)-\min\Big(a_{2,1}(\rho), \\
\nonumber
& \! a_{6,1}(\rho,\mu_1) \! + \! a_{3,2}(\rho,\mu_1), a_{1,1} \! + \! a_{3,2}(\rho,\mu_1) \! + \! a_{4,2}(\rho,\mu_1) \Big), \\
\label{EqdeltaR1}\\
\nonumber
\delta_{R_2}& \triangleq \min\Big(\kappa_{1,2}(\rho),\kappa_{2,2}(\rho),\kappa_{3,2}(\rho)\Big)-\min\Big(a_{2,2}(\rho),\\
\nonumber
&  \! a_{3,1}(\rho,\mu_2) \! + \! a_{6,2}(\rho,\mu_2), a_{3,1}(\rho,\mu_2) \! + \! a_{4,1}(\rho,\mu_2) \! + \! a_{1,2}\Big), \\
\label{EqdeltaR2}
\end{IEEEeqnarray} 
\begin{IEEEeqnarray} {ll}
\nonumber
\delta_{2R}& \triangleq \min\Big(\kappa_{4}(\rho),\kappa_{5}(\rho),\kappa_{6}(\rho)\Big)-\min\Big(a_{2,1}(\rho)+a_{1,2},  \\
\nonumber
& a_{1,1}+a_{2,2}(\rho), \\
\nonumber
& a_{3,1}(\rho,\mu_2)+a_{1,1}+a_{3,2}(\rho,\mu_1)+a_{7,2}(\rho,\mu_1,\mu_2), \\
\nonumber
& a_{3,1}(\rho,\mu_2)+a_{5,1}(\rho,\mu_2)+a_{3,2}(\rho,\mu_1)+a_{5,2}(\rho,\mu_1),  \\
\label{Eqdelta2R}
&  a_{3,1}(\rho,\mu_2)+a_{7,1}(\rho,\mu_1,\mu_2)+a_{3,2}(\rho,\mu_1)+a_{1,2}\Big),  \qquad \\
\nonumber
\delta_{3R_1}& \triangleq \kappa_{7,1}(\rho)-\min\Big(a_{2,1}(\rho)+a_{1,1}+a_{3,2}(\rho,\mu_1)\\
\nonumber
& +a_{7,2}(\rho,\mu_1,\mu_2), a_{3,1}(\rho,\mu_2)+a_{1,1}+a_{7,1}(\rho,\mu_1,\mu_2)\\
\nonumber
& +2a_{3,2}(\rho,\mu_1)+a_{5,2}(\rho,\mu_1), a_{2,1}(\rho)+a_{1,1}+a_{3,2}(\rho,\mu_1)\\
\label{Eqdelta3R1}
& +a_{5,2}(\rho,\mu_1)\Big), \\
\nonumber
\delta_{3R_2}& \triangleq \kappa_{7,2}(\rho)-\min\Big(a_{3,1}(\rho,\mu_2)+a_{5,1}(\rho,\mu_2)+a_{2,2}(\rho) \\
\nonumber
& +a_{1,2}, a_{3,1}(\rho,\mu_2)+a_{7,1}(\rho,\mu_1,\mu_2)+a_{2,2}(\rho)+a_{1,2}, \\
\nonumber
& 2a_{3,1}(\rho,\mu_2)+a_{5,1}(\rho,\mu_2)+a_{3,2}(\rho,\mu_1)+a_{1,2}\\
\label{Eqdelta3R2}
&+a_{7,2}(\rho,\mu_1,\mu_2)\Big). 
\end{IEEEeqnarray} 
\end{subequations}
Note that $\delta_{R_1}$ and $\delta_{R_2}$ represent the gap between the active achievable single-rate bound and the active converse single-rate bound; $\delta_{2R}$ represents the gap between the active achievable sum-rate bound and the active converse sum-rate bound; and, $\delta_{3R_1}$ and $\delta_{3R_2}$ represent the gap between the active achievable weighted sum-rate bound and the active converse weighted sum-rate bound. 

\noindent
Finally, it is important to highlight that, as suggested in \cite{SyQuoc-TIT-2015, Suh-TIT-2011}, and \cite{Etkin-TIT-2008}, the gap between $\agicnof$ and $\cgicnof$ can be calculated more precisely. However, the choice in \eqref{Eqdelta} eases the calculations at the expense of less precision. 

\noindent
Figure~\ref{FigGapGICNOF} presents the exact gap existing between the achievable region $\agicnof$ and the converse region $\cgicnof$ for the case in which ${\overrightarrow{\SNR}_1=\overrightarrow{\SNR}_2=\overrightarrow{\SNR}}$, ${\INR_{12}=\INR_{21}=\INR}$, and ${\overleftarrow{\SNR}_1=\overleftarrow{\SNR}_2=\overleftarrow{\SNR}}$ as a function of $\alpha=\frac{\log \INR}{\log{\overrightarrow{\SNR}}}$ and $\beta=\frac{\log\overleftarrow{\SNR}}{\log\overrightarrow{\SNR}}$. Note that in this case, the maximum gap is $1.1$ bits per channel use and occurs when $\alpha=1.05$ and $\beta=1.2$.

\section{Conclusions}

An achievable region and a converse region for the two-user G-IC-NOF have been introduced. It has been shown that these regions approximate the capacity region of the two-user G-IC-NOF to within $4.4$ bits per channel use. 

\balance

\bibliographystyle{IEEEtran}
\bibliography{IT-GT}

\begin{thebibliography}{1}
\providecommand{\url}[1]{#1}
\csname url@samestyle\endcsname
\providecommand{\newblock}{\relax}
\providecommand{\bibinfo}[2]{#2}
\providecommand{\BIBentrySTDinterwordspacing}{\spaceskip=0pt\relax}
\providecommand{\BIBentryALTinterwordstretchfactor}{4}
\providecommand{\BIBentryALTinterwordspacing}{\spaceskip=\fontdimen2\font plus
\BIBentryALTinterwordstretchfactor\fontdimen3\font minus
  \fontdimen4\font\relax}
\providecommand{\BIBforeignlanguage}[2]{{%
\expandafter\ifx\csname l@#1\endcsname\relax
\typeout{** WARNING: IEEEtran.bst: No hyphenation pattern has been}%
\typeout{** loaded for the language `#1'. Using the pattern for}%
\typeout{** the default language instead.}%
\else
\language=\csname l@#1\endcsname
\fi
#2}}
\providecommand{\BIBdecl}{\relax}
\BIBdecl

\bibitem{QPEG-TR-2016-2}
V.~Quintero, S.~M. Perlaza, I.~Esnaola, and J.-M. Gorce, ``Approximate capacity
  of the two-user {G}aussian interference channel with noisy channel-output
  feedback,'' INRIA Grenoble - Rh{\^o}ne-Alpes, Lyon, France, Tech. Rep. 8861,
  Mar. 2016.

\bibitem{SyQuoc-TIT-2015}
S.-Q. Le, R.~Tandon, M.~Motani, and H.~V. Poor, ``Approximate capacity region
  for the symmetric {G}aussian interference channel with noisy feedback,''
  \emph{IEEE Trans. Inf. Theory}, vol.~61, no.~7, pp. 3737--3762, Jul. 2015.

\bibitem{Tuninetti-ISIT-2007}
D.~Tuninetti, ``On interference channel with generalized feedback ({IFC-GF}),''
  in \emph{Proc. of International Symposium on Information Theory (ISIT)},
  Nice, France, Jun. 2007, pp. 2661--2665.

\bibitem{Suh-TIT-2011}
C.~Suh and D.~N.~C. Tse, ``Feedback capacity of the {G}aussian interference
  channel to within 2 bits,'' \emph{IEEE Trans. Inf. Theory}, vol.~57, no.~5,
  pp. 2667--2685, May. 2011.

\bibitem{Etkin-TIT-2008}
R.~H. Etkin, D.~N.~C. Tse, and W.~Hua, ``Gaussian interference channel capacity
  to within one bit,'' \emph{IEEE Trans. Inf. Theory}, vol.~54, no.~12, pp.
  5534--5562, Dec. 2008.

\end{thebibliography}
\balance

\end{document}